# INNOVATION ÉDUCATIVE EN SCIENCES DE L'INFORMATION

## *Une expérience en sciences marines*

ENRIQUE WULFF BARREIRO

Le modèle 2.0 d'innovation éducative revient, en ce qui concerne ses modes de développement, à un usage et une production des livres de texte et de la personalisation des notes de classe. La controverse, nécessaire, autour de la delégation de la connaissance socialement construite à une autorité centrale unique vs l'octroi aux individus de la connaissance additionelle dont ils ont besoin pour participer leur plans à des autres, répondrait au dynamisme actuel de l'enseignement à distance. Pour introduire a ces stratégies de travail avec l'information scientifique en sciences marines, un cours en ligne s'est vu transformer en une occasion de vivre et d'évaluer l'accès ouvert.

Concerning its development in the virtual classroom, the web 2.0 educational innovation means the use and the production of textbooks and the personalisation of the classnotes. The controversy, that is a precondition of awareness, organized around the assignment of knowledge to a central authority vs its grant to individuals who need it to share their plans with others, would meet the present dynamism of e-learning. To introduce these training strategies with scientific information in marine sciences, an online course was transformed into an opportunity for evaluating and living open access.

## Introduction

Le modèle 2.0 d'innovation éducative revient en ce qui concerne ses modes de développement, à une discussion au sujet de la controverse 'Bazzar vs Cathédrale' et à un usage et une production des livres de texte et de la personnalisation des notes de classe. Alors, la controverse, nécessaire, autour de la délégation de la connaissance socialement construite à une autorité centrale unique vs l'octroi aux individus de la connaissance additionnelle dont ils ont besoin pour participer leurs plans à d'autres répondrait au dynamisme actuel des économistes anti-schumpetériens. (Hayek, 1945) Quant au second sujet, la production de supports – textes à l'aide – spécialisés de communication éducative, il puise ses sources sur l'établissement d'unités d'organisation qui aient en vue la production de dépôts de sa "mémoire" artificielle soit archives, registres, bibliothèques, systèmes de continuité. (Simon, 1947). L'action rationnelle de choix des moyens aux prises avec la consécution des fins éducatives, est aux remontrances économicistes du fait des insuffisances de provision des marchés donnant sur les échos du 'droit à vivre' (Polanyi, 1963) – ici 'droit à l'éducation'. La création continuelle de connaissances – chère aux technologies de l'information, dont les stratégies de mise à jour savent jouer du caractère extensionnelle de la récupération de l'information moyennant la miniaturisation temporelle des dispositifs communicationnels d'emmagasinement – sait gré à une genèse des systèmes nationaux d'éducation (Bernal, 1939).

Le titre de cet article, *innovation éducative en sciences de l'information : une expérience en sciences marines*, a pour point de référence deux styles d'ingénierie de software fondamentalement différents. Celui en place dès la mise en marche en 2002 de la web 2.0, assimilé aux courants innovateurs de la e-education et débiteurs des techniques de recherche télématique avancées des contenus audiovisuels par raisonnement sémantique. Et celui qui mettant en vue les interets de la "science privée" reconnait de premier abord les interets propietaires des données et connaissances en circulation y constraignant l'accès. Les systèmes d'aide à la prise de décision pour cadres seniors en sciences marines peuvent s'identifier à ces facteurs de succès.

Du point de vue de la médiation communicationnelle en pédagogie, ces deux façons d'agir en ce qui concerne la conception de l'outillage se retrouvent aussi bien, dans la façon de concevoir et de déployer les applications pour mieux développer et maintenir ses fonctionnalités (Bisognin *et al.*, 2007). Ainsi, d'une part, les aspects prescriptifs de la conception de l'enseignement seraient le sujet de la ingénierie pédagogique. Est-ce à dire le cours, ou le projet pour l'obtention curriculaire du diplôme, ou le programme d'enseignement, ou une session d'exercices, sont satisfaisants s'ils conduisent à la meilleur efficacité de l'activité pédagogique. Et, d'autre part, les technologies peuvent être l'objet – à l'image de l'aprentissage par problèmes – où le point de depart est la codification individuelle des contextes sujet de l'instruction.

Cette différence de portée s'exprime bien par l'idée de communauté. Le problème de capitalisation des objets pédagogiques, sur la base du partage d'un travail en ligne, s'accomode mal du modèle dit instructionnel lorsque l'on met en avance un code d'aprentissage par projet. Bien que l'interêt de l'innovation, en éducation à distance, pointe sur le developpement d'una pratique commune dont le but ne peut être que l'apprentissage des tâches.

En plus d'un aménagement rationnel des études, avec des mécanismes de suivi de la qualité des enseignements partagés et de la homologation internationale, le changement en cours - qui suit à la Déclaration de la Sorbonne de 1998 [1] - souhaite porter atteinte aux méthodologies éducatives (Michavila, 2009). La reflexion critique au sujet de la web 2.0, en ce qui fait à son application au domaine de l'éducation fait jouer, d'autre part, l'idée de la communication reversible (Romano, 2003). Les medias de communication auraeint des difficultés de faire pousser le but de l'élévation des niveaux éducatifs des citoyens dès par sa inscription dans les systèmes politiques.

Dans ce contexte la différentiation d'une communauté de scientifiques, de océanographes[2], et des gens dont le travail ou l'occupation donne sur des résultats de synthèse modélés et structurés par la culture de l'océanographie et son monde sensoriel, se constitue en tant que espace d'intervention non gouvernamentale (Nuere, 2000). Les bases intellectuelles de la recherche d'alliances, sur le domaine télématique, de la part d'une telle coordonée d'acteurs relèvent des besoins de ceux ayant accès à la moderne infrastructure des TIC, ainsi que à l'information relevante, mais qui n'ont pas un support d'infrastructure éducative pour permettre aux gens non pas seulement de se bénéfier de l'information mais aussi de l'habilité de créer une nouvelle connaissance pour le bénéfice de tous (Medina-Garrido *et al.*, 2006). Les différents groupes dépositaires de ces besoins et de ces usages de l'information, pointent vers la définition des audiences spécifiques. Ces circonstances et conditions spéciales sous-entendent les types d'information qu'ils produisent et en font usage en des activités diverses ainsi que l'estimation de l'importance des fluxes d'information entre et parmi les groupes considerés. Une situation spéciale et des besoins à des profils particuliers peuvent aussi bien venir des pêcheries artisanales et de l'aquaculture à petite échelle, comme d'un centre de control de la contamination océanographique. En ce sens là, la communication pédagogique ne saurait pas être encapsulée en des messages – aussi ouverts à des procès de dialogue que cela soit le cas – mais elle doit faire partie d'une communauté de pratique. Ainsi le focus n'est pas l'acquisition d'une connaissance comme quelque chose de déjà-fait, mais comme un objet à construire à l'intérieur d'une culture technologique et profesionnelle (FAO, 2009). Et à travers les échanges et les intéractions dans lesquels ils participent, soit dans le niveau doméstique, communautaire, local, regional, national ou international.

[1] "Europe is not only that of the Euro, of the banks and the economy: it must be a Europe of knowledge as well."

[2] 'Oceanógrafos Sin Frontera' est une organisation non gouvernementale sans but lucratif, avec siège en Allemagne (Kiel) et constituée en 2005. Elle serait des seules en tant que présence Web 2.0 en sciences halieutiques et aquatiques (sciences marines) (Chu et al., 2009).

Cette confluence de l'identification d'audiences en tant que communautés de pratique et de la gestion par projets d'un enseignement à distance, reprend le renouvellement de la compréhension des processus de l'apprentissage organisationnel de la part des participants en raison de ses élaborations d'une identité numérique en tant que étudiants. Identité vue comme une participation négociée dans une communauté de pratique et une trajectoire d'apprentissage. (Chanal, 2000)

### Ancrage sur le campus virtuel en Amérique Latine

Pour l'Amérique Latine l'identification d'une technologie relevant d'une transférence volontaire et imitable des connaissances et d'une bonne vélocité de diffusion, est-ce à dire de ses possibilités d'identification et de transmission, vont d'un même pas que le degré auquel les connaissances peuvent être codifiés, enseignés et observés. Ce qui est observable dans l'ancrage national des activités professionnelles d'enseignement, là où le niveau de la relation enseignant/enseigné par ordinateur émerge à partir du besoin d'une meilleure classification du contenu. La entreprise 2.0 qui va se mettre au travail[3] reprendra les fonctions de formation et d'entraînement online pour les professionels qui travaillent aux sujets des sciences marines et limnologiques. Mais avec les développements Web 2.0 se mettent en branle les avantages ajoutées de l'enchaînement dynamique entre ces fonctions. Par example, connection instantanée entre les étudiants; une haute diversité des participants atteignable à des coûts payables; flux d'information à double sense; haut standard de service dans un intervalle de temps acceptable minimisant les differences entre les liens socials que les instructeurs online et face-à-face établissent avec leurs étudiants; control de qualité des contenus; politique d'évaluation des acquis focalisée sur les interventions hebdomadaires de réalisation des travaux pratiques; et politique d'archivage des matériaux d'interchangement propres aux modèles d'étude sous examen.

Pour analyser les processus de capture et de catégorisation des contenus, la reestructuration de la documentation de façon simple et rapide pour pouvoir la présenter à chaque moment dès la perspective de l'interêt de l'étudiant, a été le critère de choix de la plataforme moodle. Son système d'étiquetage pointait vers un coût d'entretien réduit, en même temps que la souplesse de la tâche de supervision des travaux semblait être garantie, et que la confirmation de la validité des contenus était à portée. L'entourage de gestion des connaissances de la salle de cours virtuelle s'est mis au travail sur la base des coordonnées d'accès européennes[4], bien qu'en espagnol et sur la base de garantir une solution technologique optimale du problème horaire pour assurer la connection efficiente entre les deux continents. L'empreinte de cette politique de captation et traslation envers les étudiants en tant qu'usagers avancés du potentiel chercheur en accès libre, mobilise leur ancrage national autour de l'interoperabilité de la technologie employée. Le developpement d'une ontologie de gestion des connaissances spécifique à la transférence des apprentissages occupe toujours ceux qui s'inquiètent d'une meilleure recherche et récuperation des contenus dans cet ambiance de travail en ligne en langue espagnole.

[3] AqCen, au Pérou, par l'entremise de son portail de formation online Aquacampus.

[4] Ce qui n'est pas sans rappeler le besoin de relations inter organisationnelles entre une ONG allemande – impliquée en premier chef dans la transaction pédagogique -, un campus scientifique espagnol qui effectivement pilote la virtualisation des contenus, et un campus virtuel péruvien où tous les participants doivent connecter.

Pour cette répresentation des façons de faire et des façons d'être en situation de éducation à distance des deux cotés de l'océan Atlantique, une charactéristique importante serait la collation extensive de dispositifs communicationnels longuements distincts. Moodle assure que ceux mis en marche par le projet d'action pédagogique se familiarisent avec les localisations et les contenus des leçons. Le facteur d'autonomie de l'apprentissage-enseignement est venu par la mise en route des fonctionnalités de conférence web d'un système disponible en accès libre, dimdim.

La synchronisation pour la construction des identités numériques des participants, pour la capture des expériences éducatives dans des contextes événementiels donne sur le repérage spatial et temporel de l'ambiance de l'apprentissage en ligne.

*Description des participants qui assistent au cours*

Les participants assistants au cours, en provenance du Pérou, de la Colombie, du Chili, de l'Equator, du Brésil, du Venezuela, de Cuba , de l'Uruguay, du Méxique, de l'Allemagne et de l'Espagne. L'offre su tours a eu un retour positive de la part de Cuba sans que finalement une inscription se soit ensuivie. Le professeur fût facilité par invitation dès l'Espagne. Le responsable informatique de gestion du cours a servi de support dès le Pérou.

Pour contraster ces usagers de la salle de cours virtuelle, ils ont été divisés en trois panneaux principaux.

*Panel des institutions : (a) centres de données océanographiques, (b) instituts de recherche en sciences marines associés à des universités publiques ou institutions de l'état; (c) bibliothèques universitaires; (d) organisations non gouvernamentales; (e) doctorants (biologie marine, océanographie, sciences agraires); (f) professionels indépendants (technologie de la pêche).*

Ces institutions offrent un eventail des besoins en education, qualifications professionelles et activités académiques et spécialisées dans le champ de la médiation technologique et de la consultance en gestion des connaissances des participants. La façon dont elles sont organisées, quelle est son organigrame, comment elles fonctionnent, et quel est son personnel, où sont elles localisées (sur une base centralisée vs. décentralisée; en des compagnies privées ou sur des organismes publiques) donnent une première expression aux attributs de la communication institutionnelle en jeu. Les conditions de travail et les produits de ces vecteurs de transmission des connaissances font référence à la génération d'un signal, d'une guide précise, à l'égard de l'accesibilité numérique, par considération des coûts, c'est à dire en tant que possibilité de mesurer le temps et l'effort que ceux en besoin des connaissances peuvent 'payer' pour l'obtenir et des typologies des retours pédagogiques qu'ils attendent. Une question se pose à propos des tendances futures de l'actif essentiel que pour les récepteurs de ces transfers suppose l'assimilation par apprentissage et expérimentation personnelle des connaissances non codifiés ou partiellement codifiées. De telles distinctions importent de par la decisión du détenteur de réveler ou garder sécrète la connaissance la plus récente sur les recherches encore en stade embryonnaire (Pénin, 2003).

*Panel des 'experts': (a) spécialiste en information océanographique en charge de la tutorie virtuelle; (b) le gestionnaire du réseau télématique.*

Pour révéler à un public les compétences de la gestion corporative publique/privée en termes d'externalités, sous forme de connaissances libres et gratuites, il faut s'ajuster à la dimension restreinte de la divulgation. Les conséquences pour les enseignants qui évoluent sur le Web 2.0 derivent de l'espace social de l'innovation en sciences marines dès l'interieur duquel ils bénéficient des échanges de connaissances internes propres à ce reseau. Son travail en tant que vecteur de transmission, où sont envisagés des effets positifs liés à la réputation d'innovateur, est un moyen de faciliter l'entrée dans des reseaux de production des connaissances, de repérer des partenaires compétents et potentiels (et ainsi de promouvoir la formation d'alliances inter-organisationnelles), et de resoudre des problèmes d'asymétrie d'information.

*(a) Spécialiste en information océanographique en charge de la tutorie virtuelle*

Le choix de la part du groupe d'océanographes d'un agent particulier en tant que professeur est le choix pour une localisation géographique favorable dès l'Espagne envers l'Amérique Latine. Ce qu'incite un agent à participer à de telle projet tiendrait aux effets favorables du partage des connaissances (et de rabais de coopérer à la diffusion gratuite et à l'accès libre) et à l'effet de taille qui permet d'influencer les registres communitaires d'innovation sous la forme d'externalités positives. Ses tâches en tant que responsable de l'architecture du cours incluent les critères de développement de l'application au-délà des travails de sa mise en marche et entretien. C'est à dire, il finit les négotiations d'inscription des étudiants (autrement gérés par le responsible télématique en ce qui concerne les frais d'accès et la remise de l'information indispensable pour la première connections), il nomme les conventions pour les bases de données et de connaissances utilisés en tant que recours éducatifs (les forums, dispositif de chat, initialisation des protocoles d'accès par web camera) et les contenus à l'interieur d'elles; et la création des formulaires et objets reusables pour la gestion du cours (échelle d'évaluation et qualification, gestion des archives des sessions d'enseignement).

*(b) le gestionnaire du réseau télématique*

Le contexte d'hétérogenéité des participants et de la localisation des dispositifs télématiques, définit le concept du rôle de gestionnaire technologique destiné à la solution des problèmes de transmission. La synchronisation étant plus particulièrement indispensable en cet échange entre les deux continents, le moment temporaire a requis l'établissement d'un signal générale de début sous forme d'une conférence virtuelle de présentation du cours. Ce qui s'est fait en décembre 2007 par l'ouverture de la salle de cours virtuelle avec l'annonce de ce cours d'inauguration des activités d'éducation à distance de la part de cette communauté d'océanographes, pour le début de l'année suivante 2008.[5] Ce travail a été assuré par le technician péruvien (chercheur en aquaculture au département de biologie, microbiologie et biotechnologie de l'Université Nacional del Santa), par l'entremise de la distribution (par courrier électronique) des invitations à participer à une reunión Web, produites en l'occurrence par le système des stations de travail en équipe Dimdim. Aussi bien, la directrice du campus virtuelle est venue sur les écrans des participants (le 20 février 2008) pour donner la bienvenue lors de la première session du cours. Elle a assuré aussi la gestion des huit étudiants boursiers[6].

[5] http://www.aquahoy.com/index.php?option=com_content&task=view&id=2993&Itemid=1

Un des panélistes, à la tâche définie d'étudiant par ailleurs, a su assurer, d'autre part, la communication dès l'Europe (Allemagne). Au potentiel pour fournir une communication richement diverse s'est uni sa capacité pour réperer des enjeux d'entraînement dans l'usage des systèmes électroniques tout au long du cours.

L'acheminement et bonne resolution des problèmes d'instrumentation informatique ont été facilités par la mise en disposition de la part de ce trio de spécialistes, de ses addresses 'skype' et de ses téléphones portables et fixes. Outre le carnet d'adresses électroniques s'est vu encourager ainsi le contact personne à personne où porter l'emphasis sur l'émergence des besoins d'instruction au potentiel non entamé de la part des participants aussi bien que du point de vue du système de communication mis en place. La migration à un systeme internet 2.0, comme celui de la flexible culture de l'innovation 'skype', a engagé une prestation additionnelle d'avancement à l'usage des participants.

*Panel des étudiants*

Ils peuvent être distribués selon trois niveaux interconnectés d'autonomie, c'est-à-dire, selon sa capacité d'analyser et resoudre ses problèmes d'apprentissage par transposition optimale sur ses respectives entourage de travail. Il s'agit dans le premier niveau de ceux dans le jour par jour des résultats de gestion, le second concernerait les gens en contact avec des arrangements institutionnels, et le focus du troisième rapporterait sur la construction des images, des valeurs, principes et critères pour guider la politique de pêche et de recherche marine sur un chemin consistent (Suárez de Vivero J.L. *et al.*, 2008).

Dans cette distinction et sur le premier ordre, il apparaît que la diversité du comportement local des pechêries et de la connaissance ecologique de la part de ceux dont le moyen de vie est la pêche determine une identification d'un 40% des étudiants avec l'activité des problèmes quotidiens de l'information produite et demandé. Ce sont les gens pour lesquels l'éducation à distance online puise sur le protagonisme et le sens de la responsabilité de la part des étudiants. Ils attendent de l'outil télématique la forme et le mechanisme qui va possibiliter la monitorisation constante de l'interaction avec le professeur et le groupe. Il s'agit de collaborateurs dans l'effort de recherche (tel que des ingénieurs des pêches, ou des bibliothécaires).

---

[6] Ils ont vu le règlement des frais d'immatriculation réduits de moitié (25$)).

Les systèmes de normes, lois, régulations, procedures et organisations occupent les visées pédagogiques de ceux à insertion profesionelle dans le secteur éducatif. Bien que en plus d'une occasion il n'aient pas de l'expérience en médiation technologique de la provision des connaissances externes techniquement de succès pour ses organisations. Ils se montrent à la faveur des hypothèses qui les ont conduit à suivre cet enseignement parce que il va pouvoir leur faire comprendre comment les scientifiques incorporent les données de ses efforts à l'intérieur de ses modèles, autant qu'a faire le point sur les difficultés d'intelligence des sources et le degré auquel les préoccupations des pecheurs sont prises en compte. Des professeurs d'universités et des étudiants de thèse concernés par l'interaction institutionnelle au-délà des observations ad-hoc du monde pêcheur à propos de qui pèche et quand, testent ainsi ses capacités pour se conduire d'une façon beaucoup plus critique sur la mesure de l'effort d'articulation des contenus de ses disciplines.

La recherche d'une expérience valide pour guider la politique en sciences marines (politique de contamination, et de protection des organismes marines) est la troisième position argumentale. Pas nécessairement coïncidante avec la réalisation antérieur de cours d'apprentissage à distance, ils s'inscrivent dans la lignée des étudiants aux atteintes fortement orientées vers des sujets d'interaction internationale. Les buts de distribution et alignement avec autres, d'applications de gestion des connaissances, les voient aboutir à l'assistance aux réunions des organismes internationaux de l'information marine (Conférence internationale sur les données marines et systèmes d'information, en Grèce 31 Mars-3 avril 2008). Ou la description de processus qui transforment des élements d'observation locale en chartes internationales du suivi des espèces marines, pointent vers l'approche plurinationale. Il a chez ces responsables concomitance du changement du focus d'attention des resultats locaux envers des préoccupations plus régionales et globales et une pousée des resultants de recherché à l'égard d'une intelligence plus integrée des pêcheries et de sa gestion.

L'espace de travail axé sur le regard communicationnel serait donc divisé par les resources textuelles et audiovisuelles qui focalisent le processus d'innovation et par les contextes géographiques. 19 personnes ont initialement prévue l'inscription, en provenance du Pérou (2), du Chili (2), de la Colombie (4), Venezuela (1), Cuba (1), Équateur (1), Brésil (1), Uruguay (1), Méxique (1), Allemagne (1), Espagne (3), le contact n'ayant pas pu déterminer la nationalité d'une dernière personne. Les besoins de control de la rentabilité des échanges propres à la pédagogie numérique mise en marche par la réalisation d'un projet, ont finalement conduit 9 personnes à obtenir le diplôme assurant l'action de réalisation performante.

### *Le campus virtuel Aqua Campus*

Les communications unifiées de Aqua campus, proposent pour ce cours une manière hebdomadaire de mener des conférences web. Mais, avant tout, la présentation formelle du cours de la part du professeur se tient sous la forme d'une conférence virtuelle, par invitation. L'impartition des sessions d'enseignement de la part du professeur qui exigent la présentation de diapositives Web ou le partage d'applications, sont planifiés une fois par semaine. Et d'autre part, par exemple, pour les communications quotidiennes informelles, le professeur dispose d'une interface sur son écran informatique où implementer les modèles d'ensemble du problème considéré à chaque fois, à différence de l'interface utilisateur toutefois conçue pour la génération éventuelle d'un forum. Et surtout en ligne, au long des sessions de cours et de correction des exercises et de direction des projets, le 'chat' va permettre de "converser" en temps réel. Le tableau 1 est proposé à l'intention d'exposer la méthodologie des conférences choisie.

Au miroir du système d'étiquetage de ces modes de communication au campus, la détermination des divers facteurs qui convergent sur les lignes de force des capacités en jeu, va être possible. En particulier,

– Le nombre des participants effectivement sur la matrice d'accès horaire.

– Le taux d'opérations d'assistance à la maîtrise des contenus requis par participant.

– Le type de media (video, audio, data, hybrides) qui sont disponibles pour chaque session d'impartition des contenus du cours ou de direction des projets.

– D'où viennent les étudiants, avec est-ce que cela donne de tenir à l'esprit (en l'occurrence de forme constante sur l'écran) les clés du status de celui en connection mais aussi du monde sur lequel il espère majorer ses capabilities récemment acquises.

| Typologie du scénario | Fréquence | Visibilité |
|---|---|---|
| Conférence virtuelle, resource alloué en tant que mode normal d'articulation des sessions éducatives; en complement au cours, cette fonctionalité fut appliquée aussi en décembre 2007, permettant de signaller son début. | Diffusion hebdomadaire | Tous les participants |
| Courrier électronique, les éleves doivent fournir leur addresses numériques et, celles-ci sont réconverties en addresses spécifiques de la plataforme à utilisation prioritaire en tant que dispositif du campus. | Tout au long du cours à n'importe quel moment | Le professeur et les partenaires qui s'imposent le dialogue à distance |
| Forums, disponibles et de la part du professeur et de celle des participants, en tant que moyen majeur pour la discussion en public et le partage des reéultats. | En tant que resource formelle, lisible au long des sessions une fois par semaine, et en tant que organisation informelle des conaissances distribuées et en discussion accesibles en tout instant. | Tous les participants pour les forums ouverts lors des séances du cours; et les intégrants des équipes de travail, plus le professeur, pour ceux mis en marche au sujet des projets. |
| Chat, qui met en balance les écarts de temps, de styles, et de personnalité des participants au profit des | Portant sur le crédit et la confiance de tous les participants, il est usable à tout | Les intégrants du chat d'après le formulaire de configuration, plus le professeur. |
| contenus du cours. | moment. | |
| Blog, approfondie la spécificité des réponses et des corrections des choix en place ; structure entièrement indépendante usée dans ce cours pour la réalisation du projet. | Mis en marche dès la seconde semaine. | Tous les participants |

*Tableau 1. Scénarios du campus*

Pour determiner le degré de la mediation technologique de la part des tâches en charge, il faut en determiner celles en correspondance avec les fonctions de professeur. Ainsi sur la session première, en février, il va introduire, il va donner la bienvenue, aux outils télématiques à disposition des participants. Mais avant il activera la conférence, par l'entremise de son mot de passe donnant accès à son rôle d'administrateur du resource. Les apprenants sont donc amenés à considérer ses partenaires de façon collaborative par la démarche initiale d'ouverture de la salle de cours virtuelle.

À l'instar de la séquence des événements et du flux des représentations des divers scénarios de conférence le professeur authorise le premier accès des components du cours, une foit qu'il verifie qu'ils ont les privilèges pour entrer au processus en situation d'apprentissage.

L'activité d'enlargissement des signes subtils de complaisance que les participants ont su laisser voir par ses communications initiales suite à l'acquis des droits d'inscription, fourni une première pour l'acquisition concurrentielle des dimensions des points d'entrée différents en tant qu'éleves. De la part du professeur, avec l'ouverture du premier 'Forum'une enquête consacrée à estimer quelles sont les points forts et faibles de l'éducation à distance par Internet est approchée aux participants. Cet éclairage pragmatique initiale va definir l'instrument 'Forum' d'une façon remarquable, faisant en sorte qu'il inclura 56 sujets de discussions à la finition du parcours de l'apprenant en avril 2008. Sur l'initiative du gestionnaire télématique, un second forum portant sur des nouveautés (comme des documents additionels en provenance de cours postdoctorales en documentation numérique facilités par quelqu'uns des étudiants), mais surtout rappelant dès la part du professeur les délais de finition des exercises pour mieux définir la maîtrise des concepts sur chaque étape

Faisant intervenir la dimension délibérative propre des 'Blogs' le professeur initialise l'action du blog du cours par l'évaluation de l'état du niveau des connaissances des apprennants sur les problèmes structuraux des processus de la publication en accès libre. Il s'agit d'une enquête pour fixer le sujet du projet du cours. La tension dialectique, ancrée sur l'expérience vecue des individus au travail, que cette condition didactique pour le succès au cours impose, permet au professeur d'introduire sa vision de l'apprentissage comme une notion de frontière entre la pratique professionelle et la constellation communicative définie par la platefome du campus virtuel.

Le système de qualifications est une intéressante formalisation disponible sur la plateforme dont la démarche approche l'élement confiance entre l'enseignant et ses élèves. En effet la visibilité en étant un critère premier, le nombre d'accès par ressource a une valeur pour mesurer l'activité numérique. D'autre part, l'icône pour s'engager sur le terrain des qualifications est omniprésent sur l'interface du cours, en garantie de la symétrie et de l'efficience de la procédure. En dernier lieu, l'enseignant a integré la fréquence des conférences web pour que les échanges des connaissances entre les participants s'établissent sur le niveau de la confiance mutuelle et pour qu'elles conduisent à un améliorement mesurable en honorant les engagements en ce qui concerne les dates de livraison des travaux et exercises de classe. L'argument pour l'évaluation des exercises a été de ne pas avoir retardé et/ou de les avoir en effet présentés tous. Il y a eu a remplir sept cahiers d'exercises et trois enquêtes au total, avec un délai maximum de présentation d'une semaine.pour le premiers et une considération de recevable ou pas pour les secondes. La notion d'orientation vers un objectif commun se souligne pour la qualification du projet par le suivi de cette activité sur n'importe quel des cinq scénarios en place sur le campus. Il y a eu une convergente importante sur la vision du professeur et dans l'instant de la qualification une échelle de degrés 'présentable', 'avancé', 'très avancé' s'est produite. C'est sur cette même base d'interpretation que les étudiants ont été finalement qualifiés. Et la liste des étudiants avec les pointages respectifs cumulatifs obtenus et le nombre de séances de travail résolues a été publié à la fin du cours.

L'édition des contenus didactiques a eu pour but la contribution à est-ce que l'interaction mutuelle entre les participants soit singulier à chacun d'eux. Ainsi la consécution d'une masse critique pour la mise en marche d'une communauté viable, d'une "conversation" télématique, a suivi les principes de:

– mise sur internet au campus virtuel des matériaux didactiques exposés au cours de la session hebdomadaire avant la connection effective par le service de conférence web.

– mise en place au campus virtuel des cahiers des exercises et des enquêtes immédiatement après les émissions d'éducation à distance transmises par internet.

– encouragement aux étudiants à revenir beaucoup sur l'usage maximum des resources de fouille des potentiels de la virtualité du campus (resources d'activité télématique par individu).

– édition et mise à disposition permanente des discussions soutenues en chat, questionnaires répondus, interventions suite à l'usage du blog; l'usage des références concurrentes envers les resources du campus a permis aux étudiants d'en déduire un critère de choix de ses partenaires pour l'élaboration du projet.

Comme les organisations de pechêries ont souvent exprimé ses besoins de bien écrire en science, par l'organisation d'ateliers et la publication de guides, l'interface du cours est un domaine d'expertise identitaire. L'écran apparaît divisé en trois colonnes.

La colonne de gauche offre les panneaux des informations et des actions générales du cours. Ils sont différents pour le professeur et pour les étudiants, en ce qui ressort de l'administration. Du point de vue du professeur, connecté en tant qu'administrateur électronique, la fonction fondamentale sera celle de l'activation de l'édition qui va permettre aux participants de resoudre leur problème d'être capable d'obtenir l'original des matériaux didactiques. Comme même sur web 2.0 la récupération éffective des informations de gestion des pechêries et de politique en sciences marines se base beaucoup sur la connaissance de qui a publié quoi et sur la navigation individuelle sur des pages web, la distribution (le campus numérique produit automatiquement des courriers éléctroniques à destination des participants avec la notice des nouveautés récemment mises en place sur les divers modèles d'accès des documentations du cours) et la publication rapides des contenus vise l'implementation de ces façons d'agir comme meilleur pratique dans le domaine professionnel. Les activités enregistrées sont un répertoire partagé entre tous les rôles présents dans le campus.

La colonne centrale concerne la politique éditoriale du professeur, ce sont les vincules aux contenus matériales et textuelle du tours. Ils sont publiés immediatement avant et après chaque session de conférence web éducative. Et comme les éléments extérieurs ou antérieurs à la pratique enregistrés y restent, l'aménagement de l'espace disponible porte sur la géneration continuelle des resources. L'ordre decisionnel propre à la fonction éditoriale du professeur se double d'un critère d'ouverture maximale pour l'accès aux contenus du cours de la parts des apprennants. Ainsi le diagramme hebdomadaire du cours va integrer les textes des lessons et des cahiers d'exercises, toutes les réponses à ces cahiers de la part de tous les étudiants (leur encourageant la consulte mutuelle régulière), les tutoriels sur 'Powerpoint' utilisés en cours de conférence web pour améliorer les présentations, les enquêtes et les rapports sur ces élements de control de la qualité élaborés par le professeur, le textes des chats survenus lors de chaque session, et les qualifications plus les projets du cours (dont le positionnement viendra au début de la colonne une fois le cours fini). Au cours de l'usage de ce developpement Web 2.0 l'attention a porté sur le fait que le secteur des pêcheries comporte une ample couverture de niches d'audience restreinte, ainsi s'est-il exprimé le besoin de cultiver une attitude pas entierement dépendante de la bonne volonté pour donner lieu à est-ce qu'importe à haut degré pour alimenter la réflexion éducative, c'est-à dire le partage des connaissances. L'identité des étudiants a été tenue d'être revelée à chaque instance de présentation virtuelle, ce qui a produit une atmosphère de confiance mutuelle, autrement nécessaire pour l'utilisation créative des corrections en cours d'édition des résultats finalement publiés.

La colonne de droite offre un tableau récapitulant de l'organisation temporelle de l'activité. Ainsi, les nouveautés, les prochains événements, un rapport de l'activité récente, le recensement des actualisations du cours, ou le calendrier, y trouvent ses points d'entrée.

*Description technique de l'expérience avec le système de conférence web 'Dimdim'*

La capacité prêter attention aux images émotionnelles des personnes et de leurs expressions héberge un échantillon de nouvelles initiatives dont ce cours s'est fait écho. En effet, des échelles d'observation et des interêts différents sont le reflet des différents connaissances extraites de la recherche et des pêcheries. Or la génération de l'information appropiée n'est pas suffissante, pour autant que dans le domaine des pêcheries une gestion ne peut pas être effective tant que les mesures ne sont pas considérées légitimes pas tous les participants, somme tout tant que les audiences ciblées ne sont pas identifiées, et les messages et medias à user ne sont pas coupés à mesure. C'est pourquoi une passarelle de communication entre différents discours de connaissance est nécessaire.

Sur la colonne centrale de l'interface du campus numérique, le gestionnaire télématique du campus a regulièrement mis en marche, de façon hebdomadaire, l'application software de la conférence web, pour chaque session d'éducation à distance.

En faisant usage de cette application de partage et de la largeur de bande additionnelle dont les produits spécifiques ont besoin, le professeur peut partager toutes sortes de matériaux, sa présentation (ppts, pdfs) et n'importe quelle autre ressource mise en commun (whiteboards, applications) avec les participants en diverses contrées des continents américain et européen. Le partage de la vidéo et de l'audio en simultané s'est tenu sur la base de la disponibilité de un microphone et d'une caméra Web par participant. La plateforme de travail devait supporter Internet Explorer 6.0 pour Windows 2000 ou version supérieur; ils avaient également besoin de l'installation du Flash plugin sur le navigateur.

Pour le logiciel libre à l'intention de cette tâche précise du système de conferences web utilisé, trois différents profiles ont été configurés pour modeler les diverses types de vitesse des connections des participants. Des largeurs de bande grande sur l'intranet ont été conseillées aux participants pour leur assurer une meilleure expérience à cause de sa plus haute sécurité. Suivant les instructions d'installation LAN la largeur de bande maximale de chargement est de 750 kbps; si la connection broadband est de support assuré par des systèmes de câble telles que DSL ou ISDN la spécification de largeur de bande a été de 150 – 200 kbps; quand les étudiants ont dû connecter avec le serveur du professeur la vitesse d'accès commutée a été autour des 50 kbps. La capacité du serveur a été sous les influences de la largeur de bande disponible, la mémoire disponible pour le serveur et la puissance de la CPU. Étant donné la nature de l'application - de collaboration synchronique en temps réel – la largeur de bande a été le facteur indépendant le plus important. Au total les conditions de largeur de bande ont dépendu du nombre de profils de connection différents supportés par le serveur, et du nombre de participants par conférence virtuelle.

Sur la base des profils de configuration LAN et Broadland l'application diffuse la transmission vidéo, et pour l'option Broadland avec Webcam elle supporte les 38 fps. L'option vidéo n'a pas été applicable par le moyen d'une carte d'accès à distance (dialup)

Pour les presentations web tous les participants ont pu y faire des annotations simultanément sur le support de lecture, des inscriptions sur le tableau blanc ('whiteboard'), envoyer des messages instantanés, et transmettre ses audio et vidéo, pour la plupart en temps réel. Une option pour l'invitation aux participants existe, et l'émission des invitations beta donnant sur l'accès aux conférences web a été un privilège du professeur. Le service d'hébergement du système est disponible sans frais et peut être mis en usage d'une facile manière pour des groupes de vingt ou moins personnes. (avec des interactions audio et chat).

Dès la part des participants, le cadre d'installation du logiciel a été éliminé. Pour faire cela, l'outil d'installation de l'entité 'professeur' a eu à incorporer un component ActiveX sur la base des liaisons facilitées. Et les concurrents n'ont eu besoin qu'a cliquer sur les boutons appropiés deux fois. Aucune autre configuration a été nécessaire. L'expérience complète revenant à quelques minutes à peine. En tant que logiciel libre integré sur un dispositif Moodle, la plataforme d'apprentissage en ligne peut être vue sous l'angle des LMS, bien que à différence de Blackboard la géneration des nouveaux objets ne soit pas propiétaire.

Le professeur a eu la possibilité d'incorporer les participants en tant qu'audience au même moment qu'il transfèrait les présentations (généralement sur des diapositives powerpoint, ou documents word). Ainsi la narration par diapositives a pu être transmise, suite aux facilités vidéo/audio de la version libre du logiciel des conférences Web. Entre les élèves, des problèmes de concurrence se sont associés à l'essai de transmission des photos. La qualité et l'accès ont été contingents en accord avec les conditions du reseau et les longues distances affichées.

Une fois à la salle de classe virtuelle et la session de cours débutée, la transmission vidéo avec un canal de web caméra a démarrée sur toutes les stations de travail des participants avec la visualisation in vivo de l'enseignant. Pour les étudiants l'observation directe du cours en ligne leur a permis de focaliser leur efforts sur une échelle d'interêt d'intensité maximale, d'une manière pragmatique. Néanmoins, la version du logiciel employé n'a pas permis l'archivage des versions vidéo ou audio des sessions de formation offertes.

Un aboutissement additionnel de l'identification avec les formes de discours de la web 2.0 est venue de l'option pour le chat privé, disponible pour les participants pour qu'ils s'envoient entre eux des messages privés. Ces interventions ont été un canal de communication et de collaboration performant vis-à-vis le développement des critères de multi-appartenance où puisent les tensions efficaces pour donner suite aux projets de cours. Aussi bien une option pour le chat publique a toujours été à disposition. Pourtant, en ce qui concerne l'accès audio, le partage de microphone a été limité au maximum de trois personnes. Le professeur, plus deux étudiants. Bien que tous les participants ont pu entendre les voix des trois personnes avec control audio (assigné par le professeur).

## Les espaces transversaux de l'accès ouvert

Le succès du travail virtuel avec les contenus du cours a fait suite à est-ce que l'emphasis ait porté sur les buts pédagogiques de la formation, à entraînement minimale du coté des logiciels et systèmes informatiques. Le principal facteur culturel à intensifier s'est vu exprimer de la part du temps et des energies des apprennants en ce qui tient à l'évaluation ou à l'exacte captation efficace de "l'individualisme possessive"[7]. Un bon nombre des étudiants évoluant sur des programmes de mentorat, l'approche des contenus des externalités de recherche en accès libre suivait le développement de ses attitudes et de ses pratiques quotidiennes comme une conséquence de leurs expériences professionnelles.

La plus grande partie de ce que l'on tient par une préoccupation contemporaine au sujet de l'explosion de l'information en science s'inscrit sur une conception erronée, parce que sur la base d'un modèle non valable de la nature du progrès scientifique. (Simon, 1968). La science n'avance pas par entassement de l'information; elle l'organise et la comprime. Quand l'idée de servo-mécanisme, de rétroaction, est apparue - en même temps que celle de cybernétique – il y avait des disciplines (telle la chimie organique) qui n'étaient qu'un amas de détails faiblement organisés par l'entremise de généralisations théoriques connues. La mécanique quantique et l'information quantique offrant, aujourd'hui, des puissants mohines d'organisation, bien que ces domaines de la connaissance aient augmenté ses dimensions beaucoup. Par conséquence, sans doute il est plus facile d'obtenir une position intéressante du point de vue disciplinaire convenable pour y faire des contributions originales et remarquables que par le passé lorsqu'on connaissait beaucoup moins.

[7] Cher à C.B. MacPherson.

Cette reconceptualisation en ce qui fait à l'utilisation de la revue scientifique/technique dans un système intégré à une grande échelle qui inclut aux bibliothèques, a beaucoup de conséquences sur la transition de ces revues dès copies en papier maintenues localement aux bases de données électroniques partagées. "Qui inclut aux bibliothèques... ", veut dire la fondation d'un établissement de petite taille à titre spécialisé, son développement continu sur la base d'établir des collections de référence dans les divers projets par disciplines et sa relation avec d'autres bibliothèques et bibliothécaires de la spécialité sur la base de contacts internationaux. Le monde online a un bon exemple dans le système de bibliothèques de sciences halieutiques de l'organisation des Nations Unies pour l'alimentation et l'agriculture (la FAO). Comme l'argent de le budget d'une bibliothèque seulement peut être dépensé une fois et comme au fur et à mesure que ces budgets sont plus restreints les bibliothèques ont moins d'information sur l'accès ouvert, il faut voir les tendances à la diminution de la capacité d'achat des revues en papier comme un problème de désinformation sur l'accès ouvert. (Wulff, 2008) Les coûts et les bénéfices des licences pour revues académiques dans des bibliothèques font l'objet d'une discussion tenue par les autorités et non par les bibliothécaires. Ce sont donc les auteurs qui cherchent savoir ce qu'ils doivent payer pour obtenir que son travail reçoive davantage de reconnaissance. Eux qui cherchent à formaliser l'équivalence entre le recours payant et celui qui est gratuit sur la base d'estimations de consommation/production.[8]

[8] Une fois essayées avec les manuels de contre révolte des années soixante. (Mattelart, 1992)

Le processus de production de l'information, ayant des lois de comportement dont la nature et mode d'emploi méritent de l'intérêt, quelques exemples nous sont fournis par la distribution des revenus (la Loi de Pareto), la capacité industrielle (la loi de Zipf), les articles en statistique (la distribution de Yule), les achats du consommateur (la distribution binômiale négative), la fréquence des guerres (la distribution de Poisson), la fréquence des grèves (aussi Poisson) et les oscillations dans les intentions de vote dans les élections générales. La plus complète liberté au moment de décider ou de choisir entre un ensemble d'actions, est la note commune qui définit la règle de comportement dans tous ces exemples. Ceci n'est que la sensation que nous est produite de par être en contact avec un style. Quand les circonstances seront changeantes, pour que le processus puisse être bien compris ou pour qu'il puisse être contrôlé, il faut établir un modèle satisfaisant. C'est le début d'une réduction générale de l'activité sociale à l'ordre scientifique.

Par exemple, les nombreuses innovations et les améliorations techniques dans le secteur de la pêche de hauteur fréquemment conduisent à une distorsion des mesures utilisées pour l'évaluation des paramètres. La source d'information, les statistiques de capture, ne s'avère pas toujours très fiable pour des raisons techniques et opérationnelles. Et deviennent particulièrement utiles pour l'évaluation moderne des stocks de poissons, les perçages réguliers, effectués indépendamment des opérations de pêche, avec l'aide d'équipements standard pour mesurer les variations quantitatives des ressources dans le temps et l'espace.

### *Reflexivité indéfinie : dans la relation entre les publications électroniques gratuites et les coûts des publications en papier*

La cascade de spéculations sur l'information gratuite qu'entraîne l'Internet est une condition de succès pour maximiser l'imperfection de cette information. Comme il nous rappelle l'histoire du télégraphe, la relation entre les publications électroniques gratuites et les chaque fois plus coûteuses publications scientifiques en papier, se caractérise par sa "reflexivité indéfinie", bien que des prévisions statistiques peuvent être faites. Au lieu de nous fournir des signaux clairs qui guident nôtre comportement, le système des prix sur internet fait l'objet souvent de secondes intentions, il est ambigu. C'est l'objet d'une communication systématiquement dénaturée. (Habermas, 1970) Il est dans la nature dialectique de la science que la solution apparente d'un problème révèle, en general, d'est-ce que la question correcte n'ait pas été formulée au premier abord, ou d'est-ce qu'un problème beaucoup plus difficile et intratable réside juste sous la surface qui a été triomphalement considérée en tant qu'explication. Par exemple, les réseaux biomédicaux, dans leur logique empirique, peuvent être des réseaux d'échelle libre, cela étant un des problèmes pour comprendre sa croissance dans la pratique cruciale au moment de distinguer critères de distribution de l'information dans son intérieur. Pour faire des recherches sur les bénéfices qu'un tel réseau suppose, nous devons nous demander : quel type d'accès a quel type de ressources ? Du point de vue du scientifique les ressources peuvent être des appareils, bases de données, services, collègues, ou publications. Dans l'accès il peut y avoir une variation de degré selon celui-ci soit direct, immédiatement disponible, ou libre. L'accès est direct s'il n'est pas à médiation d'une tierce partie comme des bureaux de service (ou même comme des étudiants gradués). Il est immédiatement disponible si le scientifique a accès depuis tout lieu. Il est « libre » si les coûts sont auxiliaires ou s'ils ne sont pas chargés au scientifique. Le profil standardisé de l'utilisateur scientifique d'internet serait de résulter : direct, disponible, en accès "libre" pour les collègues et avec un accès modeste aux appareils ou les services spéciaux. Des questions plus importantes tournent à propos des classes de flux d'information scientifique entre les différents canaux de communication y faisant incluer le courrier électronique, ou la manière dans laquelle la communication électronique affecte à la formation de liens interpersonnels entre des scientifiques. Il s'agit alors de trois variables qui sont dépendantes et qui correspondent à trois classes de résultats scientifiques : la publication scientifique, la reconnaissance professionnelle et l'intégration sociale. En étant deux le principales prémisses pour expliquer les bénéfices pour la science de l'utilisation du réseau. La première est une prémisse d'efficience. Les réseaux peuvent faire diminuer le temps que les scientifiques passent dans des activités improductives comme voyages à des centres d'analyse de données, ou suivi des références en bibliothèque et bases de données. D'autre part, l'utilisation du web 2.0, par le développement d'un entourage numérique personnel de formation et de aprentissage, fournit aux scientifiques l'utilisation de nouvelles techniques et l'occassion d'étudier de nouveaux problèmes. La seconde prémisse met en évidence l'accès différentiel aux ressources qui confère des bénéfices distincts aux différentes classes de scientifiques. Sur la base des critères d'avantage accumulé de ceux institutionnellement ou géographiquement situés dans le centre par contraste avec ceux qui sont dans la périphérie. Les phénomènes océanographiques requièrent des mesures in situ depuis des bateaux de recherche et commerciaux, balises, avions et satellites. Au coeur du système qui va permettre enseigner numériquement se trouvent ces critères de coordination des collections des données sur de longues distances.

Les revues d'associations professionnelles se comportent dans les bibliothèques, admettons cette métonymie référentielle, en vendant plus bon marché que les revues des éditeurs commerciaux. Elles ont une plus grande valeur scientifique, parce que centrées autour d'institutions traditionnellement prestigieuses elles jouent un rôle dominant non seulement dans l'utilisation interne de la bibliothèque mais aussi dans le prêt interbibliothécaire. Les bibliothèques comme des médiateurs dans l'information pour la pêche responsable, peuvent alors faire un effort qui résiste aux limitations dans la volonté d'autoarchivage de la part de ceux profitant de l'accès ouvert mais peu collaborateurs au maintien d'archives numériques. C'est-à-dire, les bibliothécaires pourront changer la nature de ce système en utilisant les capacités de la technologie pour fournir rapidement de l'information dans le but d'oublier le monde des abonnements, pour changer au marché libre dans l'information en science et technologie au moyen de l'approvisionnement direct de documents. Il est apparent que ce sera la division du travail qui doit porter l'initiative, au devant du monde de la gestion soit dans le cadre du développement coopératif des collections, bien dans celui des consortiums et même dans celui de la technologie.

Comme une partie de l'analyse de la croissance scientifique, le problème des revues scientifiques a des dimensions peu fréquentes dans la sphère de ce qui est humain. Pour faire une observation qui en effet l'est, la planification et la mise en oeuvre de la recherche et le maniement ultérieur des données traitées et/ou interprétées, c'est-à-dire de l'information, résulte en deux mondes en apparence très indépendants. Ceux qui travaillent avec l'information, ils considèrent leur collection, gestion et dissémination comme essentiel pour l'utilisation effective des fonds budgétaires assignés à la recherche. Et d'autre part beaucoup de chercheurs considèrent la gestion de l'information comme technique, ennuyeuse et un mal nécessaire/innécessaire. Ceci est infortuné parce qu'il y a beaucoup d'importance sociale et d'applicabilité dans le monde coloré de l'information marine et halieutique. Ainsi les travaux qui sont en rapport avec la publication des revues en sciences marines ont à voir avec la digitalisation rétrospective de documents dirigée à la confection de catalogues de publications en ligne, le contact avec les éditeurs pour leer indiquer comment et ce qu'ils doivent demander aux auteurs pour qu'ils définissent des mots clef en utilisant un thésaurus déterminé, ou la distribution de rapports pour monitorer la législation de propriété intellectuelle électronique par rapport à l'approvisionnement électronique du document.

*Un système d'accès différentiel aux prix*

La nature de l'article scientifique évolue dès la version de l'auteur (preprint), la version pour revue (arbitrée), à la version postprint (version adaptée par l'auteur). La version pour la revue peut ne pas être disponible étant donné des limitations de droit d'auteur. Par contre le preprint et le postprint en effet peuvent être archivés dans un archive électronique. Ce qui est essentiel est l'optimisation du coût de la publication et la création d'un système d'accès différentiel aux prix. Ce qui les instituts de recherche pourraient résoudre à un coût imperceptible. En effet les institutions de recherche pourraient publier les résultats de leur recherche dans leurs propres pages web et dans un format d'accès ouvert. Obtenir les choses pour son compte a quelque chose d'irrésistible mais les bourses ou les fonds qui soutiennent le nouveau modèle de publication d'articles peuvent s'éteindre sans que les critères pour assurer l'interopérabilité et le uniformité entre les réservoirs automatiques de publications arrivent à être créés. Sur un marché dont l'efficience dépend du caractère transversal des compétences le urgence d'une communauté professionnelle leader reflétera que les contenus soient intégrés horizontalement et verticalement. (Currás, 2002)

Ces deux fonctions sont en rapport avec le potentiel de développer des portails, structures éventuellement d'accès ouvert. Fonctionnellement, alors, les deux types de portails sont le horizontal et le vertical. Le portail horizontal ("hortal") a pour but un marché de masse ou de groupes d'utilisateurs sans spécifier, il essaye d'être aussi complèt et exhaustif comme possible tant thématiquement, comme dans ce qui fait au profils des utilisateurs. La classification d'information, présentation de contenus additionnels et provision de programmes de software sont aussi plus un objectif en longueur qu'en "profondité". En opposition avec les portails horizontaux, les verticales dirigent toujours ses services à des segments ou à des sous-sections définies. Cette segmentation se base sur des sujets, il se centre sur des groupes ou des portions spécifiques du marché. La limitation à des matières spécifiques conduit à la classification plus en profondeur de l'information et les services, avec le centre dans la spécialisation et par conséquent dans la qualité et non dans la quantité. Portails d'affinité, niches ou intérêt spécial sont des expressions connexes pour le portail vertical.

On dit encore qu'il dépend des auteurs eux-mêmes de faire en sorte que les chiffres actuels changent, d'après lesquels seulement autour de 20 % des articles publiés annuellement sont en accès ouvert. Alors demeure une référence est-ce que tant l'autoarchivage en accès ouvert des manuscrits, comme celui des versions admises dans des revues arbitrées bien qu'encore non publiées, font augmenter les taux de citation des articles scientifiques. Après les revues en accès ouvert et les archives numériques thématiques, ce sont maintenant les archives numériques institutionnels qui transfèrent, en tant qu'archives scientifiques ouverts pour le travail du personnel académique lui-même, les intérêts de la thésaurisation qu'assume le activité dans le secteur. Qu'est ce que la 'thésaurisation' mais un effet `bancaire' présenté comme quelque chose qui est déjà fait. Ou les auteurs investissent pour que ses travaux soient disponibles en libre accès sur un stock de publications, soit un archive numérique, soit un warehouse, ou il n'y a rien à faire. Nous avons gâté le modèle. Au contraire l'urgence avec laquelle d'autres systèmes ont été déclarés des disfonctions suffisamment graves comme pour mettre en danger l'équilibre du système, est présent dans l'histoire de 'succès' des entreprises de digitalisation de matériaux bibliographiques. Histoire qui reconnaît l'échec des livres électroniques, et qui souligne la valeur de la page à papier acide, avec le coin retourné dans l'autre sens lorsqu'il est cherché à rappeler un moment significatif de la lecture. (Schonfeld, 2003) Équilibre létal, qui est un danger pour le raisonnement. Parce que le principe d'ergodicité réside dans la base non seulement de le systématique biologique, mais dans une grande mesure dans la reconstruction générale historique aussi. C'est la sociologie qui est transformée en technologie sociale, dans la mesure où ses pratiquants se résignent. Parce qu'ils ne doivent plus rien faire, sauf écouter, voir et se taire. Ainsi des organismes internationaux dans le monde des études sur le document disparaissent, comme la Fédération Internationale de la Documentation, sans que la néfaste explication de ce fair ne dépasse la soumission plus courante devant les producteurs, 'metteurs-en-scène', et interprètes du toujours à la mode 'scénario' suivant. Lien 'scénique' dans ce cas-ci appelé 'publications en accès ouvert'. 'Iniciative', plutôt, parce que le modèle duquel il en ressort est l'organisme vif (l'hypothèse 'Gaia', par exemple) contigu à celui disponible dans la cybernétique pour laquelle la société est un système autorégulé. Non sans que la 'disfunction' qui supposait le modèle précédan Internet ait laissé d'atteindre caractère programmatique et épigonique sur des déclarations comme celle de Tokyo où le caractère choisi de l'idéal type de la distinction qui est proclamé cède le pas, lorsque l'organisme qui a ainsi induit à penser meurt, la fédération internationale de documentation et information héritier principal du mouvement documentaire européen de Paul Otlet et Henri Lafontaine, devant le « grand changement » qui suppose l'Internet et la bibliothèque "universelle" en accès ouvert. Changement dans lequel celui qui cherche va au-delà d'accéder à un document en particulier. Au lieu de cela ceux qui cherchent ils le font "en grand train", "globalement", à la recherche "de toute chose qu'il y ait par là dehors". Déréglés, les produits en concurrence ouverte vont avoir besoin de leurs propres noms sur étiquette pour rester lábiles dans le monde des grandes corporations internationaux.

Des outils de recherche et des metadata, la tournure qui pourrait conduire à renforcer le pouvoir des monopoles d'information de la part des premiers et la complexité critique de l'utilisation des metadata pour de gérer des stock d'information en accès libre, informent une certaine intuition. L'hétérogénéité des intérêts des utilisateurs, son comportement de recherche de publications gratuites, la manière dans laquelle ces dernières disposent par des rangs ses résultats, sont des règles pour prédire avec exactitude le trafic entre les publications en accès ouvert. Il s'agit des problèmes informatiques de deux technologies qui convergent. Et, par conséquent, sans qu'il y ait une intention, le web se rend plus dense par chevauchement des droits de propriété intellectuelle des différentes compagnies. Ceci peut retarder le progrès et la question par l'accès ouvert dériverait sur une recherche au sujet des potentielles réponses organisatives à cela. À l'ombre de ce renforcement de la législation antitrust il faut examiner les brevets, mais aussi la prospective du futur de la formation.

*Adopter une politique d'accès ouvert*

Pour adopter une politique d'accès ouvert nous devons tenir compte que l'information c'est toujours ce qui est informatif pour une personne. Et ce qui est informatif dépend des nécessités et des habiletés face à l'information d'un individu, bien que celles-ci soient partagées souvent entre des membres d'une même communauté scientifique. Parce que ce n'est pas tellement l'information partagée comme l'interprétation partagée, ce qui maintient aux gens réunis ensemble. L'utilisateur veut l'information dans les documents, mais le système seulement lui donne des documents. Aussi les données ont toujours une histoire, et peut-être n'aient pas un futur très certain. La connaissance des sources c'est celle de l'histoire et du futur de l'information. La valeur supposée du fait factuel peut seulement être apprécié à travers la connaissance et la signification des sources d'information. L'institutionalisation de l'analyse et de l'organisation du contenu des documents, par l'intermédiaire des archives numériques, peut produire des centres d'information qui maintiennent leurs sources d'information - c'est-à-dire, leurs documents - exclusivement pour eux-mêmes. Alors, pourquoi produire de l'information si on va l'enfermer ?

L'accès ouvert a été dénommé accès universel en tant qu'une solution au moment de structurer l'interaction complexe entre un utilisateur et un système de récupération d'information. Ce sont des options inhérentes aux catégories de dualité qui forgent des listes doubles des figures traditionnelles et de celles tendant à les remplacer. Ainsi nous parlions d'utilisateur mais aujourd'hui d'interactivité; nous disions savoir à propos d'hiérarchies, nous voyons clair aujourd'hui en termes d'adhocraties ; la traduction a été une référence, aujourd'hui l'est le multilinguisme. Si une revue a un prix parce qu'elle maximise les bénéfices de l'éditeur il faut rappeler que ce qu'on a appelé naguère domination, est ce qu'aujourd'hui on connaît par monolinguisme.

Les propositions d'accès ouvert veulent répondre à la typologie idéale implicite qui sous-entend l'interconnection entre les fichiers informatiques. Celle qui correspond à l'interconnexion authentique des idées qui plus tard peuvent se réelaborer et se redéfinir sur un réseau partagé. Résister contre ce point de vue est important parce que il s'agit d'un atavisme informatique. Qui condamne le monde de la copie en papier aux profondeurs d'un océan sur lequel il y a un mille et demi de temps réel (bien que nous devrions dire de temps passé en travaillant, en matérialisant la lecture). L'effet politique de cet idéal tend à sous-estimer les obstacles qui freinent la mise en marche efficace des archives numériques d'information en accès ouvert. En somme la domination de cet idéal partiel et simple, qui ne connaît pas son histoire, aboutit sur des interventions politiques avec des objectifs universels. Évidemment le concept d'information, outre un état mental, est aussi une décision politique.

Beaucoup de ces problèmes sont communs à la relation entre la recherche fondamentale et appliquée : comment distribuer la propriété intellectuelle, comment établir les règles qui ont rapport à la publication des résultats, quel est le calendrier des activités, quel sont les difficultés de communication. L'absence d'examen des cas d'échec, la stipulation de l'histoire de succès comme norme obligatoire pour la production de la connaissance, ne permet pas de s'occuper des propriétés qui conforment ce qu'il y a de différent dans les relations entre les personnes qui interviennent dans un et l'autre cas. Les exercices d'évaluation normative quand ils adoptent la configuration d'agence, c'est-à-dire celle d'un lien de contrats entre des individus de leur propre intérêt, ils insistent non sur le examen de leurs effets sur les moyens économiques des familles économiquement sans ressources en ce qui fait à ses membres étudiants, mais sur la conscience de l'existence et les bénéfices de l'accès ouvert. Centralité et prestige sont derrière cette dualité hiérarchie/adhocratie. Le noyau ferme d'un plan stratégique. C'est une hiérarchie de la crédibilité parce que les choses qui ne requièrent pas de l'analyse, les cas d'échec, ne sont pas incluses entre celles prévues pour déterminer le cas théorique. (Trepanier *et al.*, 2003) Comme nous voyons, ce que l'activité humaine fonctionne à posteriori, en réseau, que la caractéristique soit l'interconnection entre des fichiers, pose un problème de direction à projection sur la définition de la programmation scientifique. Quand les chercheurs accepteront comme pratiques exemplaires les cas de succès, la hiérarchie des cas croyables, ils légitiment l'intervention politique qui leur propose de y porter leer intérets, cela faisant ils perdent de l'autonomie, ils retrouvent les limites du volontarisme.

Le développement d'un réseau en accès ouvert pour soutenir une base de connaissances en technologie de pêche en ligne et en accès ouvert dans le secteur des pêcheries et les sciences aquatiques existe sur onefish, une base de connaissance au logiciel facilité par la FAO. Il est également une institution, ifremer, qui soutient l'effort coopératif des bibliothécaires dans les sciences marines pour identifier et diffuser des articles scientifiques dans un archive numérique, avano. Les règles institutionnelles d'évaluation du rendement d'une certaine manière prêtent foi à la logique du marché des obligations, de l'action intéressée, tandis que l'environnement, qui représente la véritable richesse, se dégrade rapidement. Nouvelles difficultés pour une vaste vision du caractère cumulatif et entrelacé de la recherche scientifique.

La production de la science est communiste, un bien collectif qu'illumine pour tous, comme les phares sur la haute mer et qui, comme ceux-ci, n'admet pas des bénéfices sous-litiges. Le croissance de la science coïncide avec le rejet de l'idéal du secret. L'accès ouvert c'est son état de nature.

### Apprentissage organisationnel et élaboration par projet

Absorption et usage de la connaissance à l'interieur des organisations ont été des buts à caractère général de l'effort de repérage des organigrammes et des chartes de flux qui constituent l'identité graphique de chaque participant. L'objetif primaire de l'initiative de génération des projets du cours, a été la demonstration de l'absorption des valeurs de la provision de l'information en accès ouvert. Il fait suite à l'éffort de calcul des effets logiquement imprévisibles de l'introduction d'une seconde trame d'événements d'évaluation, d'évolution temporaire différente à celle de l'assistance regulière aux sessions hebdomadaires des conférences web et à l'élaborations des cahiers d'exercises.

L'accent à été mis sur la richesse de la communication, sur la duplication aussi large que possible des nuances, de la variété, des dimensions humaines du contact personnel et professionnel. La connection des connaissances (à comprendre, à faire, à combiner), à la base de l'apprentissage à la pratique, agit en grande partie ici dès valeurs des expériences individuelles. Et l'entreprise d'élaboration des projets a eu beaucoup des intuitions où se combinaient les subtilités de la communication en ligne.

Une décision initiale clé, prise par le professeur, a porté sur son indepéndance fonctionnelle des critères de choix en jeu en ce qui concerne la conformation des groupes de travail. Moyennant la difusión des contenus du cours, et le travail de classe sur la salle virtuelle de classe en ses différents scénarios, le regroupement idoine pour apporter leur contribution au projet s'est fait jour. Aussi, l'absence intentionnelle de control sur ce sujet a été utile pour préciser que les projets tenaient aux aspects de communication, changement de modèle des services (de négoce), et de comportement corporatif, ainsi qu'à les technologies de l'accès ouvert en information en sciences marines et halieutiques.

Le programme de développement des projets a cherché à intéresser les participants, sous la rubrique entraînement, à: l'invention des formes de coopération pour l'engagement des formes de développement des stratégies promouvant l'accès ouvert; à l'interprétation des enjeux, objectifs, compétences et risques y adjoints; et à l'accroisement de leur motivation par la prise de connaissance des ressources déjà en pratique dans le cadre de son utilisation et cherchant à capitaliser les démarches d'accélération du temps qui s'ensuivent. C'est donc une considération en tant que 'joueurs' qui prend le pas au cours de l'autonomisation des apprennants praticiens de singularisation dans la conduite de leur projets. La communication entre les membres des équipes, en faisant usage du campus virtuel (mais aussi des adresses skype, de courrier électronique, numéros de téléphone portable et des fax à utilisation déterminante sur le registre d'évitation des risques entraînant des pertes de contenu), s'est livrée à démonstrer en temps réel la valeur du système de l'accès ouvert en tant qu'outil pour le travail en collaboration et l'échange des connaissances. La discussion entre la logique de changement propre au temps pour élaborer le project et celle orientée envers l'absortion des connaissances a su maintenir le focus sur le but général (aussi bien suggéré par la métaphore de l'entraînement) d'encourager les membres des équipes des projets à découvrir le potentiel de leur organisations. L'accent mis sur le contact personne-à-personne, les besoins humains, et les conditions de mise en marche et action des systèmes technologiques et des archives numériques des connaissances.

Une première enquête pour fixer le sujet du projet a été distribué dès le début aux participants. Les huit questions formulées ont tenu sujet à propos: des difficultés majeures à l'encontre des initiatives d'accès ouvert; des opportunités sensibles aux interêts du partage de l'information; des nouveaux services à obtenir/faciliter en raison de cette collaboration; des mesures de succès en l'occurrence; des contextes et matières où les étudiants seraient plus à l'aise pour garantir l'élaboration du projet; de leur orientation professionnelle (océanographie, pêcheries, technologies de l'information, bibliothèque, moyens de communication); années d'expérience (à intervalles de 5 ans, jusqu'en plus de 20); et à propos de leur expérience en archives numériques, moissonneurs, metadata, propiété intellectuelle, et accès ouvert.

Les résultats de l'enquête ont pointé vers des intérêts particulièrement portants en ce qui fait à l'analyse de la qualité des entrepôts numériques, l'évaluation des projets et des ressources sur le chemin des organigrammes de travail, le diagnostique des normatives d'accès libre et des politiques. Ceux des apprenants avec un acquis d'expériences antérieures sur le domaine de l'information numérique en accès libre, ont su associer leur conscience des difficultés à celles, non triviales, du degré d'évolution des paniers technologiques disponibles. En effet il paraît qu'une diffusion des connaissances est sans effets négatifs lorsqu'une avancée technologique importante peut être diagnostiquée de la part de l'institution ou organisme qui décide de mettre en accès ouvert ses matériaux de recherche. D'aucuns ont dû mettre en rapport la question de savoir si les documents mis volontairement en accès libre, sous le régime d'utilisation des logiciels libres, restent utilisables en accès libre (en particulier en ce qui concerne l'emploi des sous-routines de modélisation numérique). Les personnes en contact avec les autorités maritimes ont mis l'accent sur des problèmes d'archivistique numérique par rapport aux documents oubliés à l'état de conservation mauvais.

Un second document, 'structure du flux de travail d'une archive numérique', a été fourni en même temps, qui suggère l'extension recommandée du projet (50 pages), les formats (adobe, word), plus un schéma avec des questions à répondre pour servir de point d'ancrage au projet. En ce carnet de travail figuraient les adresses électroniques du moissonneur de l'Ifremer, du portail pour l'information océanographique de l'Amérique Latine et de la Caraïbe de la commission océanographique intergouvernementale de l'unesco, et du portail du réseau pour le développement de l'information océanographique pour l'Amérique latine et les Caraïbes (odincarsa) du comité de travail pour l'échange international des données océanographiques (iode). Par le biais de ce questionnaire un choix multiple a été recherché de la part des engagés dans le développement de cet apprentissage. Quatre questions sur les produits, formats, infrastructure et publique de l'accès ouvert, précédant une requête en matière de définitions concernant les standards, les outils, et les rôles; au sujet des structures d'emmagasinement; du maniement des auteurs en ce qui concerne la sélection et approbation, conversion et stockage de ses matériaux; et les mécanismes d'accès (par index, charte, route de navigation, tables des contenus, et critères de simplicité et vitesse). Référence est faite au portail waicent de la Fao, pour ce qui fait au logiciel de gestion de l'information électronique (EIMS).

Le succès du travail en équipe virtuel en quatre des cinq groupes ayant pris part en l'élaboration des projets, s'est vu prouvé par le volume de l'usage du campus virtuel, l'enthusiasme des participants, et la prompte amélioration de l'utilisation de l'environnement numérique avec le temps. Ainsi, les stratégies d'intervention des étudiants par rapport au professeur sur le chat des sessions, se sont établies sur la règle 80/20 dès la quatrième session de transmission (le vendredi, 19 mars 2008). Quand la logique positive de collaboration à l'intérieur d'un groupe a trouvé des difficultés, il faut évoquer des raisons de collusion des profiles en déça des expériences d'identification et satisfaction des besoins des pêcheurs, cultivateurs des poissons, vendeurs et négociants. Dans ce cas là, la consolidation de l'information aux prises avec le choix du motif pour le project est finalement venue par délégation du professeur.

Un remarquable épisode de membres d'un équipe principalement intéressés à l'échange des données, et auxquels le potentiel de provision de communication richement variée occupait moins, a révélé la valeur de la vision en ligne concurrentielle sur un scénario de conférence web. En ce type de situation virtuelle, la connection personne-à-personne confère une valeur aux vecteurs ouverts de transmission des connaissances beaucoup plus expéditive que la tentative d'abordage des experts pour leur faire délivrer les données pertinentes. De la sorte cette forme d'acquérir des droits sur la distribution d'une solution à interpréter au cas par cas, affiche sa manque d'éfficacité vis-à-vis les problèmes récurrents. Dès lors le développement d'un archive numérique des solutions des problèmes fréquemment rencontrés détient l'interêt des participants et leur fait voir les bénéfices d'un mode d'exploitation en accès ouvert des résultats de leur recherche.

La technologie n'elimine pas le besoin des réunions personnelles, elle privilégie une relation de proximité par petit entourage. La mobilisation des conditions pour la pratique partagée entre 15 personnes situées à 14 localisations géographiques différentes, entre l'Amérique et l'Europe, résulte en ce que les participants peuvent découvrir des rapports surprenants et usables entre leur engagements. Des suggestions à points de départ différents peuvent aider à résoudre un problème difficile. Les conversations télématiques peuvent simplement donner aux participants un meilleur sens d'est-ce qui se fait dans d'autres parts. L'ouverture à l'inattendu a été un des principes opératifs du travail visant à la réalisation des projets originaux en tant que point de référence, parce que l'innovation créative cherche l'imprévu par définition.

Le développement de cette information anecdotique, en provenance des annotations locales et distantes des apprennants, a su être important. Jusqu'un certain degré les différents groupes se sont classés par leur différente appréciation d'est-ce qui constitue la validité d'un fait. En reconnaissance du rôle de l'imprévisible pour conduire une culture de la collaboration à la prise des décisions à l'intérieur des organisations. Ces différences pourraient être vues comme plus formelles que substantives, mais en tant que réponses basiquement positives pour la configuration des équipes de travail, intégrent et interprètent les responsabilités liés à l'élaboration des projets de cours. D'autre part, elles suggèrent de plus, que des problèmes d'identité institutionnelle, courent ici à une profondeur majeure par rapport aux problèmes spécifiques à la culture. Des intéractions à petite échelle modélées en cours de classe virtuelle par la marche des projets, à l'issue des procédures d'information investissant les apprennants sur leur lieu de travail, définissent les critères de composition des groupes.

Sans réfuter les démarches relevant des cultures de la connaissance (logique de sa gestion et logique du changement aux prises avec les contextes organisationnels des décisions) la quête d'une masse critique pour la constitution des groupes de projet a conduit des administrateurs et des coordinateurs de systèmes d'information, sur deux centres de traitement des données, en tant que chef d'équipe. Est-ce qui a généré un impact sur les projets dont la chronologie et le contenu se sont vus atteints. Les facteurs qui ont influencé ces deux prises de contrôle viennent du fait des climats de travail, de la différente intelligence des fonctionnalités ad hoc, et des stratégies pour la diffusion universelle en accès ouvert.

D'une part, le projet développé autour du centre de contrôle de la contamination du Pacifique sur la côte colombienne, prend à part un travail de nature militaire. Le climat de agilité et créativité du centre ne favorise pas comme évident est-ce qu'un bien éventuellement commercialisable soit mis en circulation en dehors des conditions de monopole d'exploitation. Remarquable exemple de capabilité pour le bon déroulement d'une structure multilocal, le projet propose l'extension pour qu'elle puisse être utilisée n'importe où dans le centre (la 'fédération des actifs') d'une meilleure et locale innovation en ce qui concerne l'échange des données océanographiques. Obtenue dans le cadre du programme du comité de travail sur l'échange international des données océanographiques de la commission océanographique intergouvernementale *de l'*unesco, une telle initiative est promue sur un logiciel au code source disponible. Cela signifie qu'une condition principale pour que le régime de libre accès soit à portée, est la diffusion volontaire et gratuite des connaissances de la part de ceux qui participent à des projets de logiciel libre. L'autorité maritime et de recherche du centre consciente de ce qu'elle peut gagner en compétitivité par le dépassement de la dichotomie entre opérationalité et affaires militaires, le projet traduit les besoins d'efficacité et d'innovation en termes de consécution d'un but corporatif. Il souligne que gagner en compétences est le résultat de l'actif en train de constitution, une nouvelle infrastructure pour la recherche. Cet avantage concurrentiel sousentend le choix d'un modèle pour la production libre des livres et journaux élaborés par le centre de recherche de l'institution. Sous un régime d'arbitrage de la part d'un comité éditorial. Les services portuaires de sûreté à Tumaco, Buenaventura, Guapi et Bahía Solano, la base navale du Pacifique, les garde-côtes du Pacifique, le comité régional pour la prévention et l'attention aux désastres (crepad), pechêries du Pacifique colombien ainsi comme la communauté académique et scientifique d'ordre national et international seraient interessés par cette production de ce centre. L'administratice des technologies de l'information du centre s'est vu accompagnée d'un océanographe brésilien.

D'autre part, la certitude de la stabilité de la connaissance au sein des organisations enchaîne dans un second projet à l'expérience singulière d'un lieu géographique et d'un milieu naturel propre aux systèmes d'information environnementale marine. En Colombie un tel système porte le nom de siam, et il a été étudié par une seconde équipe de travail. Aux interêts pour la cartographie marine, les paramètres de l'houle océanique, les évidences des courants océaniques et les évaluations de leurs vitesses, s'est ajouté la cartographie du suivi des ressources des pêcheries. Dès points de vue issus de l'ingénierie de la océanographie et de la biologie marine une approche à l'institut des recherches marines et côtières, sur la côte de l'Atlantique, a été conduite facilitée par le coordinateur des systèmes d'information de l'institut. Le projet fait partie de la culture corporative de l'institut et milite pour une expression évoluée de la valeur des techniques numériques en accès ouvert. À la faveur d'une plus grande receptivité de l'histoire de succès relative aux systèmes d'information géographique, l'application de l'analyse spatiale, et l'élaboration et mise en circulation des cartographies digitales des écosystèmes marins et côtiers. Les caractéristiques des archives numériques en place ont été passé révue par rapport à la navigation, richesse des objets de connaissance transferrés, vitesse de dissemination et de reception de la connaissance, critères d'absorption, perte d'originalité et usage. Quelques collections numériques à caractère bibliographique ont aussi fait l'objet d'une exposition (articles du bulletin de l'institut et livres sur un logiciel unesco). Outre l'ingénieur du cadastre sur place à l'institut, deux biologistes marins, méxicain et colombien, en formation en tant que doctorants.ont constitué l'effectif du projet.

Les échanges du troisième groupe ont donné suite à un projet à propos des techniques et outils pour la conception et l'entretien d'un entrepôt de données numériques à la direction nationale des ressources hydrauliques (dinara) de l'Uruguay. Une recherche principalement descriptive d'une architecture de bases de données relationnelles. En effet l'apprentissage organisationnel lié à l'estimation de réduction des temps et des coûts est l'element important pour un parcours tranversal à travers les projets de cet organisme. Cela a su inclure une exposition des données et indicateurs (les matrices de décision) obtenues en provenance des bases de données opérationnelles à partir des processus d'intégration et control de la qualité des données. Ainsi que son emmagasinage en des structures à accès efficace pour les consultations avec le logiciel de traitement analytique en ligne olap, qui satisfont partiellement des besoins outils pour la prise des décisions. Une comparaison des stratégies de traitement des données opérationnelles du programme des pêcheries du gouvernement et de celles de l'entrepôt dinara est offerte. Les données biotiques, abiotiques ainsi que celles à caractère historique, relèvent aujourd'hui d'une résolution automatique des inconsistences, à chaque fois qu'elles sont déposées dans l'entrepôt numérique. Le travail de la bibliothécaire du centre de documentation de la direction nationale dans le déroulement de ce projet, s'est enrichi des questionnements spécifiques d'un océanographe allemand et d'un ingénieur de la pêche venezolien. Cette description sert à mettre en valeur les capacités prospectives de ce projet quant aux conditions d'adaptation de cette institution, dont les performances actuelles incluent aujourd'hui un archive numérique en accès ouvert, à contenu bibliographique.

Dans le tissu des interactions et des actions propres à l'approche institutionnelle de l'archivage numérique, la quatrième équipe de travail a adopté le point de vue de l'apprentissage basé sur l'exploration de nouvelles solutions. Le déploiement d'un environnement numérique de travail à l'institut national d'océanographie en Inde ayant retenu leur attention. Initialement pour se poser des questions à propos du comment évaluer un tel dévelopment, et aussi à propos du degré de validité des prédictions qui eussent pu s'ensuivre. Ainsi sont approchés l'elément de change sous forme de l'enquête à laquelle fut soumis l'institut en 2004 pour les pétitions des documents, et l'élément de stabilité résultat en 2006 de la décision de prendre en main la mise en exécution d'un archive numérique. La crédibilité de la solution est passée sous examen avec l'idée guide d'est-ce qu'aujourd'hui il est plus important que par le passé de tirer autant de valeur que possible de la connaissance organisationnelle. Une reconnaissance explicite en tant qu'actif corporatif est avancée pour cet archive numérique. L'énumération des critères d'emmagasinement et utilisation a réussi à faire comprendre est-ce que sa gestion et l'inversion que cela impose, sont faites avec la même attention prêtée pour obtenir de la valeur des autres actifs plus tangibles à l'institut. Une bibliothécaire espagnole su montrer toute la complexité de ce dispositif, avec l'engagement d'un biologiste espagnol.

Au premier chef mettant sous analyse l'archive numérique 'aquatic commons' qu'embrasse les environnements marins, estuariens et d'eau douce, le sujet du cinquième projet considère aussi le moissoneur donnant accès ouvert aux ressources électroniques associées aux sciences marines et aquatiques, 'avano'. Le projet réalise une articulation entre le niveau individuel de consulte et le niveau organisationel avec quelques détails au sujet du logiciel. La numérisation étant éffectuée par des organismes de type documentaire (centres de données et bibliothèques scientifiques et universitaires, associations professionnelles) il en ressorte la médiatisation d'un fonds souvent fort peu connu. L'identification de ces outils a été l'œuvre d'une bibliothécaire péruvienne.

## Conclusion

La conception du Web 2.0 comme modèle de l'enseignement en ligne a promu une active participation des étudiants dans un cours virtuel enlaçant l'Amérique Latine et l'Europe. Assumant ses responsabilités dans les processus d'apprentissage, et dans les dynamiques innovatrices qu'on prétendait relancer, des contenus rélatifs à la gestion en accès libre des publications en sciences marines leur ont permis de signaler et échanger ses connaissances.

L'instauration d'un cadre télématique de conférences web hebdomadaires entre les deux côtes de l'Atlantique, et quelques villes du Pacifique américain significatives pour la recherche en sciences marines, a déterminé les conditions de l'infrastructure, en faisant inclure les outils de matériel et logiciel informatique. En tan qu'espace d'apprentissage cette formation à distance à eu a accepter des configurations horaires à diagrammes de dispersion incontestable. Pour fixer une politique d'impartition permettant d'atteindre tous les participants les scenarios d'accès, consulte et demonstration se sont varies autant que la technologie du service du campus virtuel l'a permis.

Pour assister les responsables pédagogiques et téchniques à la conduite des contacts quotidiens et des sessions de classe hebdomadaire, une enquête sur les habilités initiales au maniement des effets automatiques et aux savoir faire à propos de l'initiative de l'accès ouvert s'est distribué dès le début. Plus tard, l'importance de l'implication des participants sur le domaine des données océanographiques, dans ses environements nationaux et régionaux, a fait de la manipulation et analyse des principaux types et formats d'archives numériques l'incitation principale à suivre, mise en place pour la consécution des buts des corpus thématiques.

L'accès ouvert a été introduit à partir d'un idée de la recherche documentaire des connaissances aux besoins de réponse complexes. La tâche d'identification des connaissances portant sur l'identité des valeurs pour la prise des meilleures décisions. Les reputations internes aux organizations en ce qui fait aux process de reperage pour l'échange de valeur numérique, prennent une allure nouvelle avec la conception de l'initiative pour l'accès ouvert. Les apprennants ayant un rôle tant interne comme externe différent ont assuré le profil de ses rôles indépendants au cours des discussions enregistrées et disponibles sur l'interface du cours.

La Web 2.0 est conçue pour ce rang de functions assurant la connection dinamyque entre les groupes de travail, avec le professeur, les horaires standards, les coûts abordables de transmission, le support professionnel équivalent à celui qui pourrait être preté en direct par un opérateur humain.

Les difficultés en la consécution de la publication scientifique sont confrontées avec les meilleures opportunités que le mouvement pour l'accès ouvert en offre. Cependant, des critères de revision par les pairs scientifiques, de qualité des sources d'information, d'évaluation des archives numériques, des normatives et spécifications quant à la propiété intelectuelle viennent rencontrer les participants, en ce que les défis de la gestion de la connaissance assurent plus le partage de la connaissance que l'encombrement des connaissances.

À l'enrichessement mutuel entre les pratiques sociales différentes des apprennants, s'est agrée la présentation synthétique d'un projet. La structure et le contenu de ce mode de garantir, en tant que pratique collective, la suffissance envers les contenus du cours a fait l'occassion d'accomplir l'éxpérience satisfaisante d'un travail en équipe sur les ancrages théoriques et les concepts de pratique développés pour construire le lien avec les notions apprisses en salle de classe virtuelle.

Remerciements

Maria de los Angeles Ginori Gilkes. Acuario Nacional de Cuba. Ciudad de La Habana, Cuba. Deborah Cohen. Marylhurst University. Oregon, USA.

## Bibliographie